\documentclass[11pt,english,a4paper]{article}

\usepackage[utf8]{inputenc}
\usepackage[T1]{fontenc}
\usepackage{babel}
\usepackage{graphicx}
\usepackage{amsmath,amssymb,slashed,bbm,xcolor}
\DeclareMathAlphabet{\boldmathe}{T1}{cmr}{bx}{it}
\def\mtxt#1{\quad\hbox{{#1}}\quad}

\def\E{\mathrm{e}}
\def\D{\mathrm{d}}
\def\tr{\mathrm{Tr\,}}
\def\ha{\frac{1}{2}}
\def\Z{\mathbbm{ Z}}

\def\beq{\begin{eqnarray}}
\def\eeq{\end{eqnarray}}

\def\tr{\,\mbox{tr}\,}

\def\Tr{\,\mbox{Tr}\,}

\def\be{\beta}

\def\la{\lambda}

\def\pa{\partial}

\def\ph{\varphi}

\def\Ga{\Gamma}
\def\De{\Delta}
\def\La{\Lambda}
\def\Si{\Sigma}

\begin{document}
\begin{center}

\textbf{On the functional renormalization group for the scalar
field on curved background with non-minimal interaction}
\vskip 4mm
{
Ilya L. Shapiro$^{a,b,c,}$\footnote{Email address: shapiro@fisica.ufjf.br},
\
Poliane de Morais Teixeira$^{b,d,}$\footnote{
Email address: poliane@fisica.ufjf.br},
\
Andreas Wipf~$^{e,}$\footnote{Email address: wipf@tpi.uni-jena.de}
}
\end{center}
\vskip 4mm
\begin{center}
{\sl
(a) \ D\'epartement de Physique Th\'eorique and Center for Astroparticle
Physics,\\
Universit\'e de Gen\`eve,24 quai Ansermet, CH–1211 Gen\'eve 4, Switzerland
\vskip 2mm

(b) \ Departamento de F\'{\i}sica, ICE, Universidade Federal de Juiz de Fora
\\
Campus Universit\'{a}rio - Juiz de Fora, 36036-330, MG, Brazil
\vskip 2mm

(c) \ Tomsk State Pedagogical University and Tomsk State
University, Tomsk, Russia
\vskip 2mm

(d) \ SISSA, Via Bonomea, 265, 34136, Trieste, Italy
\vskip 2mm

(e) \ Friedrich-Schiller-Universität,
Theoretisch-Physikalisches-Institut,\\
Max-Wien-Platz 1,07743, Jena, Germany
}
\end{center}
%%%%%%%%%%%%%%%%%%%%%%%%%%%%%%%%%%%%%%%%%%%%%%%%%%%%%%%%%%%%%%%%%%%%%%%
\vskip 4mm

\begin{quotation}
\noindent
\textbf{Abstract.} \
The running of the non-minimal parameter $\xi$ of the interaction
of the real scalar field and scalar curvature is explored within
the non-perturbative setting of the functional renormalization
group (RG). We establish the RG flow in curved space-time in the
scalar field sector, in particular derive an equation for the
non-minimal parameter. The RG trajectory is numerically explored
for different sets of initial data.
\vskip 3mm

%%%%%%%%%%%%%%%%%%%%%%%%%%%%%%%%%%%%%%%%%%%%%%%%%%%%%%%%%%%%%%
\vskip 2mm

\vskip 2mm
\noindent
%%%%%%%%%%%%%%%%%%%%%%%%%%%%%%%%%%%%%%%%%%%%%%%%%%%%%%%%%%%%%%%%
\end{quotation}

%%%%%%%%%%%%%%%%%%%%%%%%%%%%%%%%%%%%%%%%%%%%%%%%%%%%%%%%%%%%%%%%%
%%%%%%%%%%%%%%%%%%%%%%%%%%%%%%%%%%%%%%%%%%%%%%%%%%%%%%%%%%%%%%%%%
%%%%%%%%%%%%%%%%%%%%%%%%%%%%%%%%%%%%%%%%%%%%%%%%%%%%%%%%%%%%%%%%%
\section{Introduction}

The renormalization structure in curved space-time is well-known
at both general and perturbative levels. In particular, it is
known that any theory which is renormalizable in flat space,
remains renormalizable in curved space \cite{book} (see also
\cite{PoImpo} for a recent review). The necessary elements of
the consistent quantum theory in curved space-time are the
purely gravitational
vacuum action, which consists of the Einstein-Hilbert term with
cosmological constant and also of the four fourth-derivative
terms. On the top of that, if the theory under discussion has
scalar fields $\ph_i$, new nonminimal terms of the form
$\,\xi_{ij} R\ph_i \ph_j$ have to be included in the action. The
renormalization group (RG) in curved space-time was introduced
in \cite{nelspan82,buch84,Toms83} (see also \cite{book}) as a
useful tool to explore the scaling properties of the theory.

The renormalization and RG
in curved space follow some important hierarchy, that means that
\ \ {\it (i)}
The RG equations for the matter fields couplings and masses do
not depend on $\,\xi_{ij}\,$ and on the parameters of the vacuum
action. More general, these equations are not affected by the
presence of external gravitational field\footnote{This is not
true if gravity is quantized \cite{BShVW}, but we do not
consider this part here.}.
\ \ {\it (ii)}
The RG equations for $\,\xi_{ij}\,$ depend on the matter
fields couplings (but not on the masses of the fields, in case
of the Minimal Subtraction scheme of renormalization), but do
not depend on the parameters of the vacuum action.
\ \ {\it (iii)}
The RG equations for the parameters of the vacuum action
may depend on the couplings (beyond one-loop approximation)
and on $\,\xi_{ij}$.

%%%%%%%%%%%%%%%%%%%%%%%%%%%%%%%%
One has to note that the running of $\,\xi_{ij}\,$ may have
some important applications, especially to inflationary models
such as Higgs inflation \cite{Shaposh-08}, because this running
is closely related to the effective potential of the Higgs field
in curved space \cite{book} (see also \cite{Markk}).
The same concerns also other
inflationary models, including the ones based on inflation,
Starobinsky inflation \cite{star} and especially its modified
version \cite{asta}. Therefore, it would be quite useful to know
whether the non-minimal parameter can experience a strong running
at some moment of the history of the universe. One of the
possibilities to observe an intensive running of $\xi$ is related
to the non-perturbative effects in the framework of the functional
RG (FRG) approach, developed by Polchinski and Wetterich
\cite{Polch-FRG}, \cite{Wetterich} (see also \cite{TimMorris}
for a similar original derivation and \cite{FRG-review} and
\cite{Wipf} for reviews and introduction to the subject). In the
present Letter we present the FRG equations in curved space, in a
background-independent covariant way similar to what has been done
before for the perturbative RG in curved space
The FRG approach on a fixed deSitter background has been previously
considered in \cite{Fr}\footnote{During the completion of this
work the related preprint \cite{Pe2015} has been published.}.
The present paper is essentially restricted to the case of a single
scalar and hence to the equation for a single parameter $\xi$.
We consider first
the local potential approximation (LPA), dealing with the most
simple theory with unbroken symmetry, and then explore the more
complicated case with the broken symmetry and wave-function
renormalization.
%%%%%%%%%%%%%%%%%%%%%
In fact, the extension of the RG flow for the broken phase is
especially interesting, because the running of the non-minimal
parameter in this case was not sufficiently well explored even
in the perturbative approach. Some potentially interesting
consequences of the RG flow for the non-minimal parameter are
related to the scale-dependence of the non-local parts of the
induced gravitational action, which emerge due to the curvature
dependence of the vacuum expectation value of the scalar field
\cite{sponta}.
%%%%%%%%%%%%%%%%%%%%%%%%%%

In the parallel work \cite{FlowXi} we will also
consider generalizations like the theory with a more general form
$f(\phi)R$ of non-minimal interaction at quantum level and in different
dimensions.

The paper is organized as follows. In Sect. 2 we describe the
general scheme of FRG in curved space-time and especially
derivation of the RG equation for $\xi$. Sect. 3 is devoted
to the numerical analysis of the equation for $\xi$.
Sect. 4 describes the FRG in the scalar theory with broken
symmetry. Finally, in Sect. 5 we draw our conclusions.

%%%%%%%%%%%%%%%%%%%%%%%%%%%%%%%%%%%%%%%%%%%%%%%%%%%%%%%%%%%%%%%%%%
%%%%%%%%%%%%%%%%%%%%%%%%%%%%%%%%%%%%%%%%%%%%%%%%%%%%%%%%%%%%%%%%%%
%%%%%%%%%%%%%%%%%%%%%%%%%%%%%%%%%%%%%%%%%%%%%%%%%%%%%%%%%%%%%%%%%%
\section{FRG for scalar field with nonminimal coupling}
\label{sect2}

The renormalizable theory of a single scalar $\phi$ in
a curved space starts from the classical action of the form
\beq
S &=& \int\limits_x \,\Big[-\frac{1}{2}\phi\triangle_g\phi
+\frac{\xi}{2}R\phi^2+V(\phi)\Big]
\,+\, S^\mathrm{grav}[g]\,,
\label{act1}
\eeq
where we assumed Euclidean signature and use the notation
$\int_x \equiv \int d^4 x\sqrt{g(x)}$. Furthermore,
$S^\mathrm{grav}[g]$ corresponds to the vacuum action as
described in the Introduction. We expect to discuss the FRG
flow for the vacuum part in a separate article, so it will
not be seriously dealt with in the present Letter or in
\cite{FlowXi}. $V(\phi)$ is a classical potential, which
may be restricted to the form $(1/4!)\la\phi^4$ in case we
intend to remain within the scope of perturbatively
renormalizable theories.

As usual in the RG approach, at quantum level all quantities
start to depend on the scale which we identify as $k$. The
practical use of the FRG approach implies the choice of
truncation scheme, which we choose in a most simple way,
assuming that the effective average action is
\beq
\Ga_k
&=&
\int\limits_x
\Big[-\frac{Z_k}{2}\phi\triangle_g\phi
+\frac{\xi_k}{2}R\phi^2+u_k(\phi)\Big]
\,+\,
\Ga_k^\mathrm{grav}[g]\,.
\label{act3}
\eeq
This truncation includes a scale-dependent effective potential $u_k$,
a wave function renormalization $Z_k$ and the running nonminimal
parameter $\xi_k$ which does not depend on the momenta or on the
field $\phi$. The invariant cutoff action has the form
\beq
\De S_k
&=&
\frac12 \int\limits_x \,\phi R_k(-\triangle_g)\phi\,,
\mtxt{where}
R_k(-\triangle_g)=Z_k r_k(-\triangle_g)\,.
\label{act5}
\eeq
$R_k$ is assumed to have the well-known properties
of a cutoff function \cite{FRG-review}.
The anomalous dimension is defined as
\beq
\eta_k
&=&
-\frac{k\partial_k Z_k}{Z_k}=-\frac{\partial_t Z_k}{Z_k},
\mtxt{where}
t=\log\frac{k}{\mu}\,.
\label{act7}
\eeq
When the scale $k$ runs from the UV-cutoff $\Lambda$ to the IR,
the dimensionless scale parameter $t$ runs from
$\log(\Lambda/\mu)$ to $\,-\infty$.

The Wetterich equation for the scale dependent effective average
action reads \cite{Wetterich,TimMorris},
\beq
\partial_t \Gamma_k[\phi]=\ha \Tr \left(
\frac{\partial_t R_k}{\Ga^{(2)}_k[\phi]+R_k}\right)\,,
\label{wetterich}
\eeq
where $\Ga^{(2)}_k$ indicates a second variational derivative
with respect to the scalar field and $\Tr$ includes the coincidence
limit and covariant integration over the space-time variables.
For the truncation (\ref{act3}) the {l.h.s.} becomes
\beq
\partial_t\Gamma_k
&=&
\int\limits_x
\Big[-\frac{\eta_k Z_k}{2}\phi\triangle_g\phi
+\frac{\partial_t\xi_k}{2}R\phi^2+\partial_t u_k(\phi)\Big]
\,+\, \partial_t\Gamma_k^\mathrm{grav}[g]\,.
\label{act9}
\eeq
In order to derive the {r.h.s.} of (\ref{wetterich}), we need
\beq
\Ga^{(2)}_k
&=&
-Z_k\triangle_g+\xi_k R+u''_k(\phi)\,,
\label{act11}
\eeq
where prime means simple derivative with respect to scalar field.
The variation of the cutoff function can be cast into the form
\begin{equation}
\partial_t R_k=Z_k\left(\partial_t r_k-\eta_k r_k\right)\,.
\label{act13}
\end{equation}
Then the {\it r.h.s.} of the flow equation
(\ref{wetterich}) takes the form
\beq
\frac12 \Tr \left(
\frac{\partial_t R_k}{\Gamma^{(2)}_k[\phi]+R_k}\right)
&=&
\frac12\Tr \left[
\frac{(\partial_t-\eta_k)r_k(-\triangle_g)}
{(-\triangle_g+r_k(-\triangle_g))
+ \xi_k Z_k^{-1} R
+ Z_k^{-1} u''_k(\phi)}\right]\,.
\label{act15}
\eeq
The equation (\ref{wetterich}) with (\ref{act9}) and (\ref{act15})
represents the covariant flow equation corresponding to the truncation
(\ref{act3}). It can be improved by including higher derivative
terms into (\ref{act9}), but then the calculations of (\ref{act15})
should also be evaluated up to the corresponding higher order of
approximation.

%%%%%%%%%%%%%%%%%%%%%%%%%%%%%%%%%%%%%%%%%%%%%%%%%%%%%%%%%%%
\subsection{Elaborating the Wetterich equation}

It proves useful to define
\beq
u_k(\phi)=\frac{m_k^2}{2}\phi^2+w_k(\phi)\,,
\mtxt{such that}
u_k''(\phi)=m_k^2+w''_k(\phi)\,.
\label{act17}
\eeq
Thus we arrive at the following form of Eq. (\ref{act15}),
\begin{equation}
\ha \Tr \Big(
\frac{\partial_t R_k}{\Gamma^{(2)}_k[\phi]+R_k}\Big)
\,=\,
\ha\Tr \Big(\frac{B_k(-\triangle_g)}{P_k(-\triangle_g)+\Sigma_k}\Big)\,,
\label{act19}
\end{equation}
where we introduced the abbreviations
\begin{align}
B_k(-\triangle_g)
& =
\left(\pa_t  - \eta_k \right)r_k(-\triangle_g)\,,
\label{act21}
\\
P_k(-\triangle_g) & =
-\triangle_g+r_k(-\triangle_g)+\frac{m_k^2}{Z_k}\,,
\label{at23}
\\
\Sigma_k(\phi,R)&=\frac{\xi_k}{Z_k} R+\frac{1}{Z_k}w''_k(\phi)\,.
\label{act25}
\end{align}
In order to analyze the flow equation in the truncation
(\ref{act3}), we need to evaluate the expression (\ref{act19})
up to the first order in scalar curvature, while the terms with
derivatives of curvature and higher powers of the curvature tensor
can be disregarded. This means that we can effectively consider an
approximation with constant $\,R$.

It is easy to note that the operators $B_k$ and $P_k$ commute.
But for a inhomogeneous field and curvature the spacetime-dependent
$\Sigma_k$ does not commute with $B_k$ and $P_k$. But they commute in
the constant curvature and constant $\phi$ approximation, the latter
corresponds to the local potential approximation (LPA).

To simplify notations in what follows we skip the arguments of
$B_k,P_k$ and $\Sigma_k$. Then the expansion of the {\it r.h.s.}
of (\ref{act19}) into a power series in $\Sigma_k$ gives
\begin{align}
&
\Tr\left(\frac{B_k}{P_k+\Sigma_k}\right)
\,=\,
\Tr \left(\frac{B_k}{P_k(1+P_k^{-1}\Sigma_k)}\right)
\,=\,
\Tr \left(B_kP^{-1}_k\frac{1}{1+P_k^{-1}\Sigma_k}\right)
\nonumber
\\
&
\hskip5mm=\Tr Q_{k,1}
-\Tr\big(Q_{k,2}\Sigma_k\big)
+\Tr\left(Q_{k,2}\Sigma_k\frac{1}{P_k}\Sigma_k\right)+O(\Sigma_k^3)\,,
\qquad
Q_{k,m}=\frac{B_k}{P_k^m}\,.
\label{act29}
\end{align}
The first term on the {\it r.h.s.} is $\phi$-independent and
contributes only to the running in the vacuum sector
$\Gamma_k^\mathrm{grav}$. Until the cutoff action is
specified, $B_k$ and $P_k$ are some unknown functions of
$\,-\triangle_g$, which should be expanded to first order
in the curvature tensor. For this end we shall apply the
useful off-diagonal heat kernel method, based on
Laplace and Mellin transforms, such that the operators in
(\ref{act29}) can be derived from the heat kernel of the
covariant Laplacian. The method is described in details in
\cite{off-diagonal} (see also \cite{FlowXi}), so here we
only sketch the main points of the derivation.

The functions $Q_{k,m}$ of the covariant Laplacian in the
Neumann series (\ref{act29}) admit representations in terms
of the inverse Laplace transform,
\beq
Q_{k,m}(-\triangle_g)
&=&
\int_0^\infty \D t\,\mathcal{L}^{-1}
[Q_{k,m}](t)\,\E^{t\triangle_g}\,,
\label{trace11}
\eeq
where
\beq
\mathcal{L}[f](s) &=& \int_0^\infty \D t\, \E^{-st}f(t)\,.
\label{trace1}
\eeq
In what follows we shall apply a useful formula \cite{off-diagonal}
%%%%%%%%%%%%%%%%%%%%%%%%%%%%%%%
\beq
\frac{1}{(4\pi)^{d/2}}\int_0^\infty \D t\,t^{-p}\mathcal{L}^{-1}[f](t)
&=&
\frac{1}{(4\pi)^{d/2}\Ga(p)}\int_0^\infty ds\, s^{p-1} f(s)\,.
\label{trace23}
\eeq
To evaluate the effective action in the given truncation
(discussed below) we need the coincidence limit of the matrix elements
$\langle x\vert Q_{k,m}\vert x^\prime\rangle$.
According to (\ref{trace11}) we may use the heat-kernel
expansion for small $t$,
\beq
\langle x \vert e^{t\triangle_g} \vert x \rangle
\,=\,\frac{1}{(4\pi t)^{d/2}}
\big[a_0(x) + ta_1(x)+
t^2 a_2(x)\,+\,\dots\big]\,
\label{trace19}
\eeq
%
%%%%%%%%%%%%%%%%%%%%%%%%%%%
to find an series expansion of $\langle x\vert Q_{k,m}\vert x\rangle$
in invariant powers of curvatures and their covariant derivatives.
The Schwinger-DeWitt coefficients $a_0,\,a_2,\,a_4,\,...$ have the form
\begin{equation}
a_0=1,\quad
a_1=\frac{1}{6}R,\quad
a_2=\frac{1}{180}\Big(R_{\mu\nu\alpha\beta}R^{\mu\nu\alpha\beta}
-R_{\mu\nu}R^{\mu\nu}+6\triangle_g R+\frac{5}{2}R^2\Big)\,,\,\dots\,.
\label{trace20}
\end{equation}
%\textcolor{red}
%{
%In what follows we restrict the presentation by the formulas for the
%traces only.
%}
In a given truncation scheme we need only $a_0$ and $a_1$, but it is not
difficult to keep also the next terms. 
Using the regulator function \cite{litim}
\begin{equation}
r_k(s) = (k^2-s)\theta(k^2-s)\,,
\label{trace13}
\end{equation}
one can arrive at the explicit expression
\beq
Q_{k,m}(s)
&=&
\frac{2k^2-(k^2-s)\eta_k}{M_k^m}
\,\theta(k^2-s),\quad M_k=k^2+\frac{m_k^2}{Z_k}\,.
\label{trace17}
\eeq
Inserting the expansion (\ref{trace19}) into (\ref{trace11}),
and using (\ref{trace23}) we obtain
\beq
\langle x\vert Q_{k,m}(-\triangle_g)\vert x\rangle
&=&
\sum_{n=0}^\infty\int_0^\infty \D t\,\mathcal{L}^{-1}[Q_{k,m}](t)
\,\frac{1}{(4\pi t)^{d/2}}\,a_n(x) t^n
\nonumber
\\
&=&
\frac{1}{(4\pi)^{d/2}}
\sum_{n=0}^\infty\frac{a_n(x)}{\Gamma(d/2-n)}
\int_0^\infty \D s\,Q_{k,m}(s)\,
s^{d/2-n-1}\,,
\label{trace25}
\eeq
where the identification $p=d/2-n$ has been used in Eq.
(\ref{trace23}).

In order to evaluate the integrals over $s$ in Eq. (\ref{trace25}),
one can note that $Q_{k,m}$ in (\ref{trace23}) are nonzero only on
the interval $[0,k^2]$, that gives
\beq
\int_0^\infty \D s\,Q_{k,m}(s)s^{d/2-n-1}
&=&
\Big(1-\frac{\eta_k}{d-2n+2}\Big)\,
\frac{1}{M_k^m}\,\frac{2 k^{d-2n+2}}{d/2-n}\,.
\label{trace27}
\eeq
After integration we arrive at the result
\beq
\langle x\vert Q_{k,m}(-\triangle_g)\vert x\rangle
&=&
\frac{2}{(4\pi)^{d/2}}\,\frac{1}{M_k^m}\,
\sum_n \Big(1-\frac{\eta_k}{d-2n+2}\Big)
 \,\frac{a_n(x)\,k^{d-2n+2}}{\Gamma(d/2-n+1)}\,.
\label{trace29}
\eeq

As we have already mentioned, for our purposes it is sufficient
to consider the $n=0,1$ terms in the last series. Expanding the
{\it r.h.s.} of the flow equation (\ref{act29}) in powers of
$\phi$ and the curvature up to the first order, in four
dimensions we meet
\beq
\ha\Tr Q_{k,1}(-\triangle_g)
&=&
\frac{1}{32\pi^2\,M_k}
\,\int\limits_x \Big[
k^6
\left(1-\frac{\eta_k}{6}\right)
+ \frac{k^4}{3}\Big(1-\frac{\eta_k}{4}\Big)\,R
+\dots\Big]\,.
\label{trace31}
\eeq
It is easy to see that these terms contribute only to the purely
gravitational terms and hence are irrelevant for the running of
$\xi$.

The second-order contribution is
\beq
\ha\Tr \big[ Q_{k,2}(-\triangle_g)\Sigma_k \big]
&=&
\frac{1}{32\pi^2\,M_k^2}
\int\limits_x
\Big[k^6\left(1-\frac{\eta_k}{6}\right)
\Sigma_k+\frac{k^4}{3}\left(1-\frac{\eta_k}{4}\right)
R \Sigma_k+\dots\Big]\,,
\label{trace33}
\eeq
were we again disregarded higher powers of the curvature.

The derivation of the third term in the expansion (\ref{act29})
requires some commutations of $\Sigma_k$ with $P_k^{-1}$, e.g.,
\begin{align}
\ha\Tr \left(Q_{k,2}\Si_k\frac{1}{P_k}\Si_k\right)
&= \ha\Tr \left(Q_{k,2}\Sigma_k\Big(\Si_k-\frac{1}{P_k}
\big[P_k,\Si_k\big]\Big)\frac{1}{P_k}\right)
\nonumber
\\
&= \ha\Tr \left(Q_{k,3}\Si_k^2\right)
-\ha\Tr \left(Q_{k,3}\Sigma_k\frac{1}{P_k}\big[P_k,\Si_k\big]
\right)\,.
\label{trace35}
\end{align}
The last term containing the commutator of $P_k$ and $\Sigma_k$ gives
rise to a running of the wave function renormalization and will be
dealt with in section \ref{sect4}. It does not contribute to the running
of $u_k$ and $\xi_k$ and thus can be neglected for the moment being.
Thus we arrive at
\beq
&& \ha\Tr \left(Q_{k,3}(-\triangle_g)\Sigma_k^2\right)
\,=\, \frac{1}{16\pi^2}\frac{1}{M_k^3}\sum_n \frac{k^{6-2n}}{\Ga(3-n)}
\left(1-\frac{\eta_k}{6-2n}\right)\,\tr(a_n\Sigma_k^2)
\nonumber
\\
&&
=\,
\frac{1}{32\pi^2\,M_k^3}
\int\limits_x
\left[
k^6\left(1-\frac{\eta_k}{6}\right)
\Sigma_k^2
+ \frac{k^4}{3}\left(1-\frac{\eta_k}{4}\right)\,R\,\Sigma^2_k+\dots
 \right]\,.
 \label{trace39}
\eeq
In the last expression we omitted most purely gravitational contributions
(not all, since for example $R\Sigma_k$ contains still a term $\propto R^2$)
and terms beyond the truncation scheme (\ref{act3}).
%%%%%%%%%%%%%%%
Similarly, one obtains for constant field and curvature
\begin{align}
&\ha\Tr\left(Q_{k,2}\Sigma_k\left(P_k^{-1}\Sigma_k\right)^{m-1}\right)=
\ha\Tr \left(Q_{k,m+1}\Sigma_k^m\right)\nonumber\\
\,
&=\frac{1}{32\pi^2\,M_k^{1+m}}
\int\limits_x
\left[
k^6\left(1-\frac{\eta_k}{6}\right)
\Sigma_k^m
+ \frac{k^4}{3}\left(1-\frac{\eta_k}{4}\right)\,R\,\Sigma^m_k+\dots
 \right]\,.\label{trace39a}
\end{align}
In what follows we ignore the purely gravitational contribution 
$\Gamma_k^\mathrm{grav}[g]$ in Eq. (\ref{act9}) and related RG flows.
We expect to consider this subject in a separate article.

%%%%%%%%%%%%%%%%%%%%%%%%%%%%%%%%%%%%%%%%%%%%%%%%%%%%%%%%%%%
\subsection{RG flow for couplings and non-minimal parameter}
Inserting (\ref{trace31})-(\ref{trace39a}) into
(\ref{act29}) and using the definition (\ref{act25}) yields
\beq
&& \ha\Tr\left(\frac{\partial_t R_k}{\Ga^{(2)}_k[\phi]+R_k}\right)
\,=\,
\frac{1}{32\pi^2M_k} \,\int\limits_x
\,\,\left[k^6
\left(1-\frac{\eta_k}{6}\right)
+ \frac{k^4}{3}\Big(1-\frac{\eta_k}{4}\Big)\,R
\right]\nonumber
\\
&\times&
\left[
1 \,-\, \frac{\xi_k R + w_k''}{Z_k\,M_k}
\,+\,
\left(\frac{\xi_k R + w_k''}{Z_k M_k}\right)^2
\,-\,
\left(\frac{\xi_k R + w_k''}{Z_k M_k}\right)^3 + \dots
\right]\label{poli5}
%% \nonumber
%% \\
%% &-&
\\
&-&\frac{1}{2Z^2_k}\Tr \left(Q_{k,3}w''_k(\phi)\frac{1}
{P_k}\big[P_k,w''_k(\phi)\big]\right)
+\dots\, ,
\nonumber
\eeq
where terms containing $R^2$ and $R^3$ go beyond the truncation scheme
and must be omitted.

%%%%%%%%%%%%%%%%%%%%%%%%%%%%%%%%%%%%%%%%%%%%%%%%%%%%%%%%%%%%%%%%%%%%%%
%%%%%%%%%%%%%%%%%%%%%%%%%%%%%%%%%%%%%%%%%%%%%%%%%%%%%%%%%%%%%%%%%%%%%%

One can assume that the classical potential for the scalar
field at the cutoff is an even function. It follows from the flow
equation (\ref{poli5}) that the scale dependent effective potential
remains even at all scales. Let us further assume that the $\Z_2$
symmetry is not spontaneously broken. Then the minimum of the
effective potential is at $\phi=0$ and we may expand $w_k$
in (\ref{act17}) as
\beq
w_k (\phi)
&=&
\sum_{n=2}^{\infty}\frac{1}{(2n)!} \, \la_{(2n)k}\, \phi^{2n}
\,=\,\frac{1}{24} \la_{4k} \phi^4 + \frac{1}{720} \la_{6k} \phi^6
+ \dots\,,
\label{poli6}
\eeq
where $\la_{(2n)k}$ are scale-dependent coefficients. Their running
is determined by the FRG equations, which will be now derived, along
with the one for $\xi_k$.

For $w_k$ in (\ref{poli6}) the last term in (\ref{poli5}) has the
form $\phi^2(\partial\phi^2+\partial^2\phi^2+\dots)$ and is beyond
the truncation (\ref{act3}). As a result the wave function
is not renormalized. Note, that for a non-even cut-off potentials
with a $\phi^3$ term, or for the flow in the broken phase, in which
the even potential $u_k$ is expanded about a non-zero mean field,
there is a wave function renormalization, as we shall see in Sect
\ref{sect4}. One can note that this situation does not depend on
the presence of curved background and can be already observed in
scalar field theories in flat spacetime \cite{FRG-review,Wipf}.

One can compare the $\phi$-dependent terms in the expression (\ref{poli5})
of equation (\ref{act9}) with (\ref{poli6}), and arrive at the truncated
flow equation for the scalar field theory nonminimally coupled to gravity,
\begin{align}
\nonumber
&-\frac{\eta_k \, Z_k}{2}\phi\triangle_g\phi
+ \frac{\partial_t\xi_k}{2}R\phi^2
+ \frac{\phi^2}{2}\,\partial_t m_k^2
+ \frac{\phi^4}{24}\, \partial_t \la_{4k}
+ \frac{\phi^6}{720}\, \partial_t \la_{6k}
\\
\nonumber
&= -\frac{1}{M_k^2} \frac{1}{32 \pi^2 Z_k} \Big[
k^6 \left(1-\frac{\eta_k}{6}\right)
\Big(\frac{1}{2} \la_{4k} \, \phi^2
+ \frac{1}{24} \la_{6k} \, \phi^4\Big)
+ \frac{k^4}{3}\left(1-\frac{\eta_k}{4}\right)
R \, \la_{4k} \, \frac{\phi^2}{2}\Big]
\\
&\hskip4mm+
\frac{1}{M_k^3} \frac{1}{32 \pi^2 Z^2_k}
\Big\{k^6 \left(1-\frac{\eta_k}{6}\right)
\Big[\xi_k R \, \la_{4k} \, \phi^2
+ \frac{1}{4} (\la_{4k})^2 \, \phi^4
+ \frac{1}{24}\la_{4k}\la_{6k}\, \phi^6 \Big]
\Big\}
\label{poli9}
\\
&\hskip4mm -
\frac{1}{M_k^4} \frac{1}{32 \pi^2 Z_k^3}
\Big\{
k^6\left(1-\frac{\eta_k}{6}\right)\frac{\la_{4k}^3}{8}\,\phi^6\Big\}
+\dots
\nonumber
\end{align}
Now, we can compare coefficients on both sides of this equation:

\begin{itemize}
\item
For the kinetic term $\phi\triangle_g\phi$ we observe that
$Z_k= \mbox{constant}$ or $\eta_k = 0$, hence there is no wave
function renormalization for an even potential in the symmetric
phase. One can fix then $Z_k=Z_\Lambda=1$ at all scales $k$.
Then, in particular, $M_k=k^2+m_k^2$, and we denote
$\,D_k=\big(k^2+m^2_k\big)^{-1}$ for the sake of convenience.

\item For the mass term the RG equation can be easily obtained from
(\ref{poli9})
\beq
\pa_t m_k^2
&=&
- \frac{1}{32\pi^2} \, k^6\, D_k^2\,\la_{4k}\,.
\label{poli12}
\eeq

\item
For the first two interaction terms we have, with $Z_k\equiv 1$,
\beq
\pa_t \la_{4k}
&=&
- \,\frac{k^6}{32 \pi^2}\, D_k^2\,
\Big(\la_{6k} - 6D_k\la_{4k}^2\Big)\,\,,
\label{poli13a}
\\
\pa_t\lambda_{6k}
&=&
-\,\frac{k^6\,D_k^2}{32\pi^2}\,
\left( \lambda_{8k}-30 D_k\lambda_{4k}\lambda_{6k}
+ 90 D_k^2\lambda^3_{4k}\right)\,.
\label{poli13b}
\eeq
In order to keep our consideration simple, we shall truncate the Taylor 
expansion for the potential by disregarding all coefficients starting
from $\,\lambda_{8,k}$, including setting $\lambda_{8,k}=0$ in
Eq. (\ref{poli13b}).

\item Finally, the non-minimal term $\phi^2R$ yields, for
$Z_k\equiv 1$,
\beq
\pa_t\xi_k
&=&
\frac{k^6 D_k^2}{16\pi^2}\,
\Big(D_k\,\xi_k - \frac{1}{6 k^2} \Big)\,\la_{4k}\,.
\label{poli11}
\eeq
%%%%%%%%%%%%%%%%%%%%%%%%%%%%%%%%%%%%%%%%%%%%%%%%%%%%%%%%%%%%%%%%%%
\end{itemize}

In a perfect agreement with the general features of the perturbative
RG in curved space-time (as described in Introduction), the equations
for the couplings $\la_{4k}$ and $\la_{6k}$ do not depend on $\xi_k$,
while the FRG equation for $\xi_k$ depends on the couplings.
At the same time, different from the minimal-subtraction
based RG, here the $\be$-functions for both couplings and
non-minimal parameter do depend on the running mass of field
$m_k$. In this respect the FRG equations resemble the
physical, momentum-subtracted based RG, developed for the
interacting scalar field in curved space-time in \cite{Bexi}, but
the mass dependence in the present FRG-equations is much stronger.

%%%%%%%%%%%%%%%%%%%%%%%%%%%%%%%%%%%%%%%%%%%%%%%%%%%%%%%%%%%%%%%%%%
%%%%%%%%%%%%%%%%%%%%%%%%%%%%%%%%%%%%%%%%%%%%%%%%%%%%%%%%%%%%%%%%%%
%%%%%%%%%%%%%%%%%%%%%%%%%%%%%%%%%%%%%%%%%%%%%%%%%%%%%%%%%%%%%%%%%%
\section{FRG flow for couplings, mass and $\xi$}
\label{sect3}

The RG equations (\ref{poli12})-(\ref{poli11}) should be
explored numerically. To this end we introduce the \emph{dimensionless}
quantities
\beq
m_t = \frac{m_k}{k}\,,\quad
D_t = \frac{1}{1 + m_{t}^2}\,,\quad
\la_{4t} = \la_{4k}\,,\quad
\la_{6t} = k^2 \la_{6k}\,,\quad
\la_{8t} = k^4 \la_{8k}\,,\,\,\dots
\label{dimless}
\eeq
which are supposed to depend on the dimensionless parameter
$\,t=\log (k/\mu)$, defined in (\ref{act7}). Then the equations
become
\beq
\pa_t m^2_t
&=&
- 2m^2_t\,-\,\frac{1}{32 \pi^2}\, D_t^2 \la_{4t}\,,
\label{num1}
\\
\pa_t \la_{4t}
&=&
- \,\frac{1}{32 \pi^2}\,D_t^2\,\big(\la_{6t} - 6 D_t \la_{4t}^2\big)\,,
\label{num2}
\\
\pa_t \la_{6t}
&=&
2 \la_{6t}\,-\,
\frac{1}{32\pi^2}\,D_t^2\,
\left( \lambda_{8t} - 30 D_t \lambda_{4t}\lambda_{6t}
+ 90 D_t^2\lambda^3_{4t}\right)\,.
\label{num3}
\eeq
Furthermore, the FRG equation for $\xi(t)$ in terms of new variable
has the form
\beq
\pa_t \xi_t
&=&
\frac{1}{16 \pi^2}\, D_t^3\,\left(\xi_t - \frac{1}{6\,D_t}\right)
\, \la_{4t}\,.
\label{numxi}
\eeq
It is easy to see that in the massless limit $D_t \to 1$ this
equation reproduce the main features of the one-loop RG equation
in the minimal-subtraction scheme, as it is known from
\cite{nelspan82,book}.

\begin{figure}
\begin{center}
\includegraphics[width=0.42\textwidth]{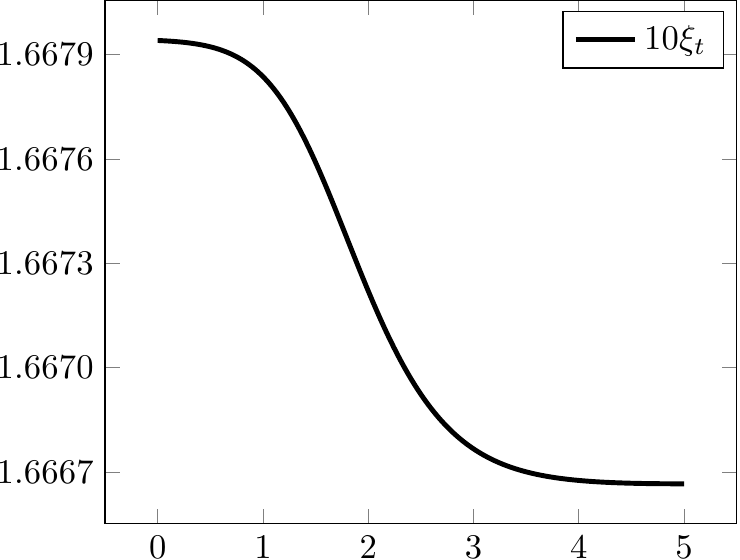} \quad
\includegraphics[width=0.40\textwidth]{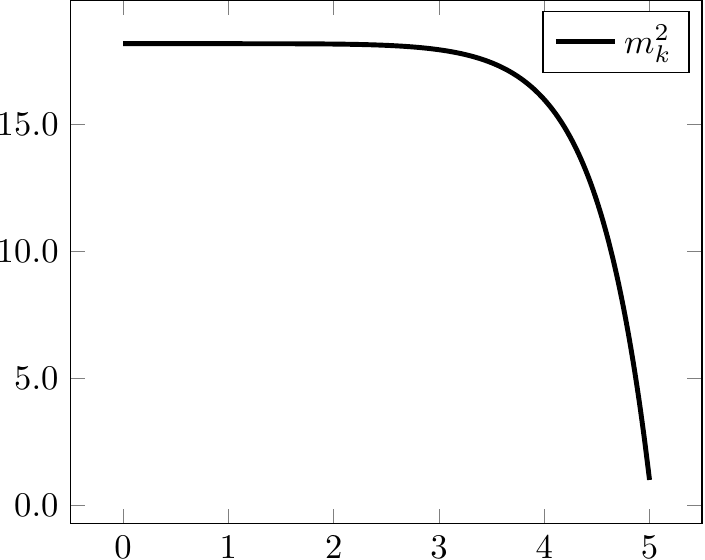}\\[3mm]
\includegraphics[width=0.42\textwidth]{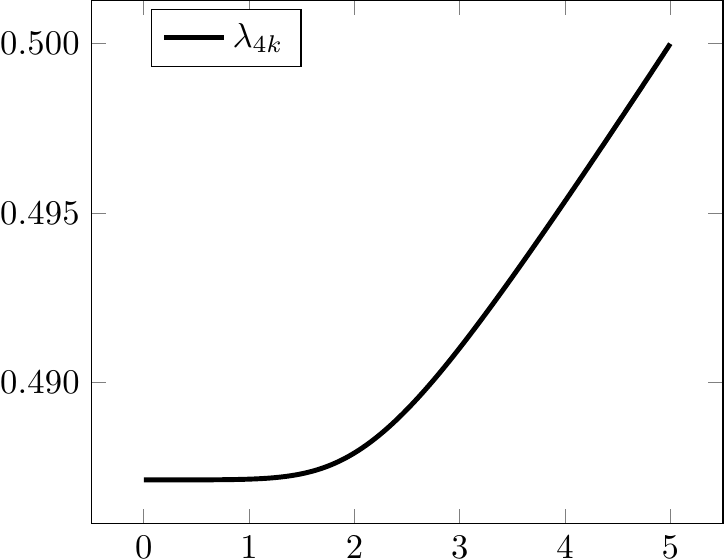} \quad
\includegraphics[width=0.40\textwidth]{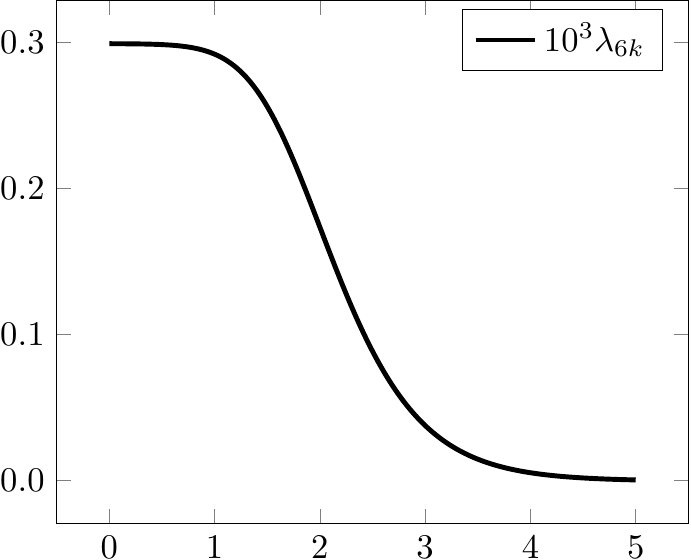}
%\begin{quotation}
\caption{\sl
Flow of the nonminimal parameter $\xi_t$ and couplings 
$m^2_{k},\,\la_{4k},\,\la_{6k}$
(in units of $\mu$) with $t=\log(k/\mu)$
in the unbroken phase. The initial data
at the cutoff $\Lambda=\mu e^5$ are $m^2=\mu^2,\;\la_{4} = 0.5$,
$\la_{6} = 0$ and $\xi= 1/6$.}
%\end{quotation}
\label{fig:parameterLPA}
\end{center}
\end{figure}

The numerical analysis of these equations shows that the RG flow
can be pretty much different from the one for the perturbative
one-loop RG running, mainly due to the mass dependence. As one
can see from the plots presented at the Figs. \ref{fig:parameterLPA},
the (dimensionful) mass grows quickly when one flows from the
cut-off scale in the UV at $t=5$ (corresponding to a cutoff value
$\Lambda=e^5\mu$) to the IR at $t=0$ which corresponds to $k=\mu$.
As we have already noted above, in the non-perturbative FRG
approach there is a sixth-power IR decoupling which is very strong
compared to the usual quadratic decoupling in the perturbative
Appelquist and Carazzone theorem \cite{AC}. As a result, in the
case under discussion one can observe that the running for all
couplings and in particular $\xi_t$ actually freezes at values
$t\lesssim 1$ or equivalently at scales $k \lesssim \mu\cdot e$.

\begin{figure}
\begin{center}
\includegraphics[width=0.4\textwidth]{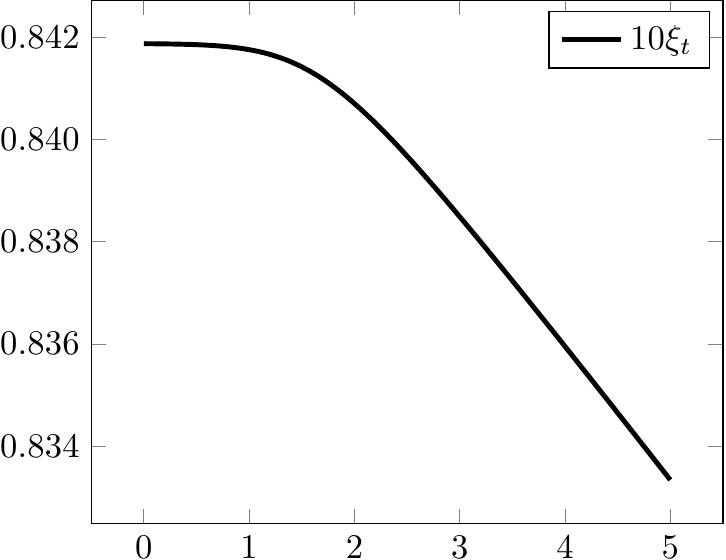} \quad
\includegraphics[width=0.40\textwidth]{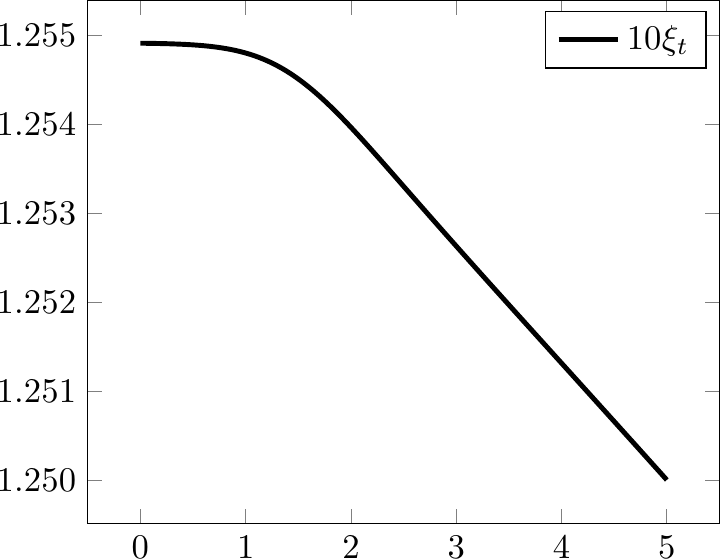}\\[3mm]
\includegraphics[width=0.4\textwidth]{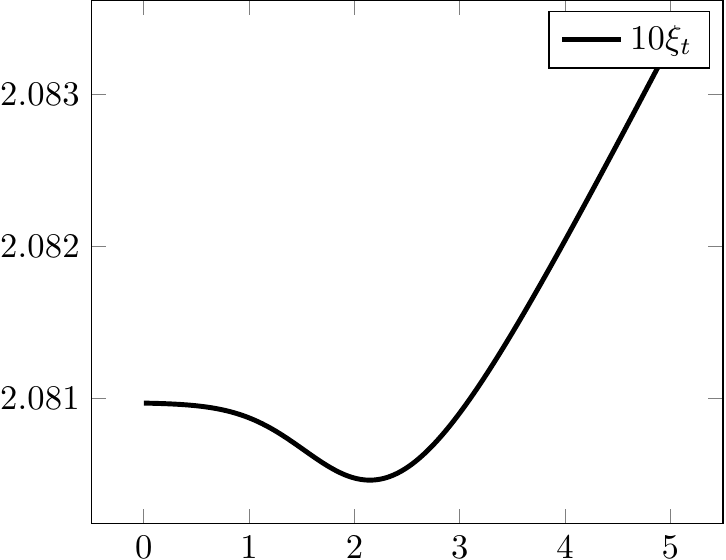} \quad
\includegraphics[width=0.40\textwidth]{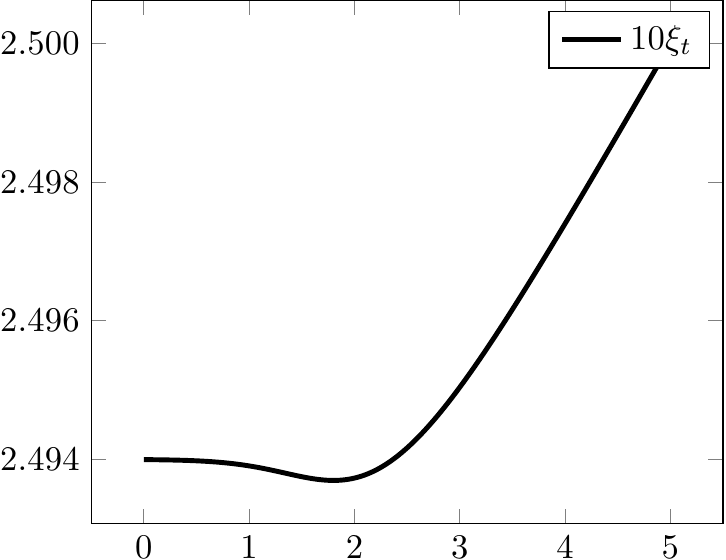}
\caption{\sl
Running of the nonminimal parameter $\xi_t$ with $t=\log(k/\mu)$
for the same initial couplings
as in Fig. \ref{fig:parameterLPA}, but with different initial values $\xi$
in the vicinity of the conformal coupling $\xi_c=1/6$. Clockwise
from top: $\xi_c-2\delta,\,\xi_c-\delta,\,\xi_c+\delta$ and 
$\xi_c+2\delta$ with $\delta=1/24$.}
\label{fig:xiLPA}
\end{center}
\end{figure}

Figure \ref{fig:xiLPA} shows the running of the nonminimal parameter
$\xi_t$ for the same initial potential as in Fig. \ref{fig:parameterLPA},
but for varying initial $\xi$-values of at the UV-cutoff. Depending on
the initial value the parameter may increase or decrease during the
flow towards the infrared. But, in all considered cases, the values
in the UV and IR  are not much different. In the flows investigated
(only some are displayed in the Figure) the relative change was only
about one percent.

%%%%%%%%%%%%%%%%%%%%%%%%%%%%%%%%%%%%%%%%%%%%%%%%%%%%%%%%%%%%%%%%%%%
%%%%%%%%%%%%%%%%%%%%%%%%%%%%%%%%%%%%%%%%%%%%%%%%%%%%%%%%%%%%%%%%%%
%%%%%%%%%%%%%%%%%%%%%%%%%%%%%%%%%%%%%%%%%%%%%%%%%%%%%%%%%%%%%%%%%%
%%%%%%%%%%%%%%%%%%%%%%%%%%%%%%%%%%%%%%%%%%%%%%%%%%%%%%%%%%%%%%%%%%
\section{Broken symmetry and FRG flow for anomalous dimension}
\label{sect4}

As we already know, in the LPA truncation
(\ref{act3}) there is no RG running for the anomalous dimension
$\eta_k$ for even potentials. At the same time, such running
is present in scalar
theory beyond one loop and it would be interesting to observe
it within the FRG approach. One of the possibilities is related
to theories with broken symmetries. This means we shall
introduce the negative mass-squared in the classical action
(\ref{act1}) and implement this information into the effective
average action (\ref{act3}) by imposing the corresponding
boundary condition at the cut-off scale $k=\La$. Then the
effective potential of the scalar field is not convex at
intermediate scales and this must be taken into account,
for in this case one has to consider oscillations near the
non-symmetric minima of this potential.

In curved space the spontaneous symmetry breaking meets
serious complications, because the position of such a minimum
is not constant for a non-constant curvature. The situation
was explored in details in \cite{sponta} and it was shown
that the non-localities emerge in such a theory even at
low orders in curvature. However, since our intention
here is to consider relatively simple cases, let us consider
the zero-order approximation and, correspondingly, assume
that the position of the minimum of the potential is homogeneous and
curvature-independent, denoted by $\phi_{0,k}$, such that
$u''_k(\phi_{0,k})=m_k^2\geq 0$ is the (physical) mass in the broken
phase. Then one has to expand the effective potential as

\beq
u_k=\la_{0k}+\sum_{n\geq 2}\frac{\la_{nk}}{n!}(\phi-\phi_{0k})^n\,,
\label{wf2}
\eeq
with small $\phi-\phi_{0k}$ and scale-dependent minimum $\phi_{0k}$.
Then
\beq
u''_k=\lambda_{2,k}+\sum_{n\geq 3}
\frac{\la_{n,k}}{(n-2)!}(\phi-\phi_{0k})^{n-2}\equiv
m^2_k+w_k''\,.\label{wf3}
\eeq
One can easily note that these definitions of $\,m_k^2\,$ and
$\,w_t\,$ are different from the previous ones, because
the expansion is performed in the spontaneously broken phase. However,
in the new notations the Eqs. (\ref{act21}) and (\ref{act25}) have
%%%%%%%%%%%%%%%%%%%%%%%%%%%%%%%%%%
almost the same form as it was before in the old
notations. But in the broken phase odd powers of $\phi-\phi_{0k}$
appear such that much more terms arise in the power series expansion
of the \emph{rhs} of the flow equation. When one calculates the
\emph{lhs} of the flow equation one must take into account that
the minimum $\phi_{0k}$ of the scale dependent potential flows.

Thus, we continue by inserting the expansions (\ref{wf2}) and
(\ref{wf3}) into the FRG equation with scale-dependent wave
function renormalization, in pretty much the same way as we did
before. Finally, changing over to dimensionless quantities
(\ref{dimless}) and to dimensionless fields according to
\begin{equation}
\chi=k^{-1}Z_k^{1/2}\phi,\quad \chi_{0k}=k^{-1}Z_k^{1/2}\phi_{0k}
\end{equation}
one can arrive at the FRG equations in the broken phase.
To simplify the notation we use the following abbreviations in what follows:
\begin{equation}
 A_t=\frac{1}{32\pi^2}\left(1-\frac{\eta_t}{6}\right),\quad
 G_n=\frac{\lambda_n}{1+m_t^2}\,.\label{brp1}
\end{equation}
The running of the (cosmological) constant $\lambda_{0t}$ is given by
\begin{equation}
\partial_t\la_{0t}+4\la_{0t}=A_tD_t\,,
\label{brp3}
\end{equation}
and since it does not feed back into the running of the remaining
couplings it will be discarded. Comparing terms linear in
$\chi-\chi_{0t}$ yields the running of the mean field,
\begin{equation}
\partial_t\chi_{0t}
+\left(1+\frac{\eta_t}{2}\right)\chi_{0t}=A_tD_t\frac{G_3}{m_t^2}\,.
\label{brp5}
\end{equation}

This flow equation ensures that $\chi_{0t}$ remains a minimum of
the scale dependent potential at all scales.
In writing the flow equations for the dimensionless couplings 
$m_t^2,\lambda_{3t},\dots,\lambda_{6t}$
in an expansion up to order $6$ we use the flow equation (\ref{brp5}) 
to simplify the resulting expressions. This way one arrives at
\begin{align}
\big(\partial_t+2-\eta_t\big)m^2_t&=
A_tD_t\Big(\frac{\lambda_{3t}}{m_t^2}G_3
-G_{4}+2G_3^2\Big)\label{m2eq}\\
\big(\partial_t+1-\frac{3}{2}\eta_t\big)\la_{3t}&=
A_tD_t\Big(\frac{\la_{3t}}{m_t^2}G_4-
G_{5}+6G_3 G_4
-6G_3^3\Big)\label{la3eq}\\
\big(\partial_t-2\eta_t\big)\la_{4t}&=
A_tD_t\Big(\frac{\la_{3t}}{m_t^2}G_5
-G_{6}+6G_4^2+8G_3G_5
-36G_3^2 G_4+24G_3^4\Big)\label{la4eq}\\
%%%%%
\big(\partial_t-1-\frac{5}{2}\eta_t\big)\la_{5t}&=
A_tD_t\Big(\frac{\la_{3t}}{m_t^2}G_6
%-G_{7}
+20G_4 G_5
+10G_3G_6-90G_3 G_4^2\label{la5eq}\\
&\hskip3cm -60G_3^2G_5
+240 G_3^3G_4
-120 G_3^5\Big)\\
%%%%%
\big(\partial_t-2-3\eta_t\big)\la_{6t}&=
A_tD_t\Big(
%\frac{G_3}{m_t^2}\lambda_{7t}-G_{8}+
20G_5^2
+30G_4 G_6-90G_4^3-360G_3G_4G_5-90 G_3^2 G_6
%+12 G_3G_7
\\
&\hskip3cm
+1080 G_3^2G_4^2
+480 G_3^3G_5-1800G_3^4G_4
+720G_3^6\Big)\,.\label{la6eq}
\end{align}
In the two last flow equations the terms containing 
$\lambda_{7t}$ and $\lambda_{8t}$ are
omitted in the sixth-order polynomial approximation.

On the top of that, in curved space one meets the new equation
for the non-minimal parameter,
\beq
\big(\pa_t - \eta_t \big) \xi_{t}
\,=\,
\frac{D_t}{16\pi^2} \, \Big[
\Big(1-\frac{\eta_t}{6}\Big)\,
\big(G_4 - 3 G^2_{3}\big)\,D_t\xi_t
\,-\,
\frac16\,\Big(1-\frac{\eta_t}{4}\Big)\,
\big(G_{4} - 2G^2_{3}\big)\Big]\,.
\label{xieq}
\eeq
In order to explore the system of equations
(\ref{m2eq})-(\ref{xieq}) we need an additional
equation which defines the scale dependence of $\eta_t$. In
order the obtain this dependence -- in our truncation it is
induced by the last term in (\ref{trace35})  or equivalently
the last term in (\ref{poli5}) -- one has to remember that the
running of all couplings and of $Z_k$ (which defines $\eta_t$)
does not depend on the presence of the curved space background.
As a result we can use the known flat-space result for
$\eta_t$ derived in \cite{ballhausen,Wipf} and recently explored
in \cite{codello2}. In terms of dimensionless
quantities the result for the anomalous dimension reads
\beq
\eta_t
%%%\,=\,\eta_k
&=&
\frac{1}{32\pi^2}\, \,\frac
{\big[ {u_t'''}(\chi_{0t})\big]^2}
{\big[1 + u''_t(\chi_{0t})\big]^4}
=\frac{1}{32\pi^2}\frac{\lambda_{3t}^2}{(1+m^2_t)^4}
\,,
\label{fleq7}
\eeq
where $\chi_{0t}$ is the scale-dependent position of the
minimum of the potential. Clearly, in the truncation
scheme considered, the renormalization of the wave function only
happens in the broken phase with non-zero coefficient $\la_{3t}$.
It follows from (\ref{brp5}) that symmetry can only be broken
if $\la_{3t}$ in the UV is non-zero, and this must be taken into
account for by a numerical analysis of the system of equations
(\ref{m2eq})-(\ref{xieq}).

\begin{figure}
\begin{center}
 \includegraphics[width=0.40\textwidth]{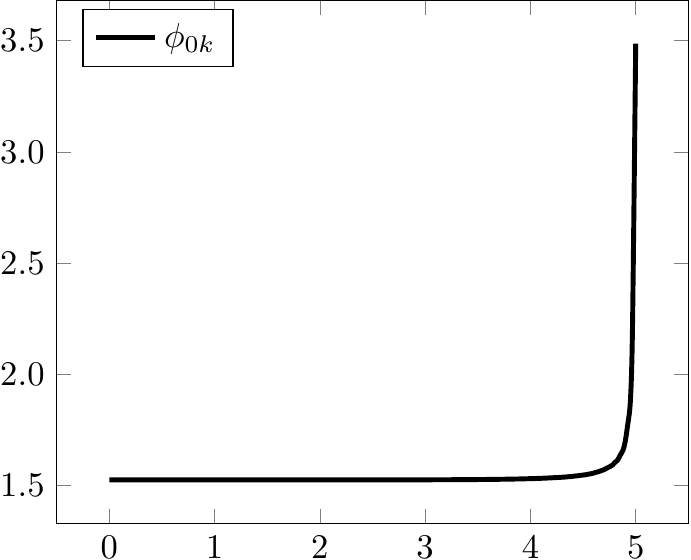}
\caption{\sl
Flow of the (dimensionful) minimum of the effective potential as
function of $t=\log(t/\mu)$.
For $\phi_{0\Lambda}\approx 3.5\mu$ (and the same initial parameters as in
Fig. \ref{fig:parameterbroken}) the system remains in the broken phase 
for all scales $k$.}
\label{fig:phibroken}
\end{center}
\end{figure}

\begin{figure}
\begin{center}
\includegraphics[width=0.41\textwidth]{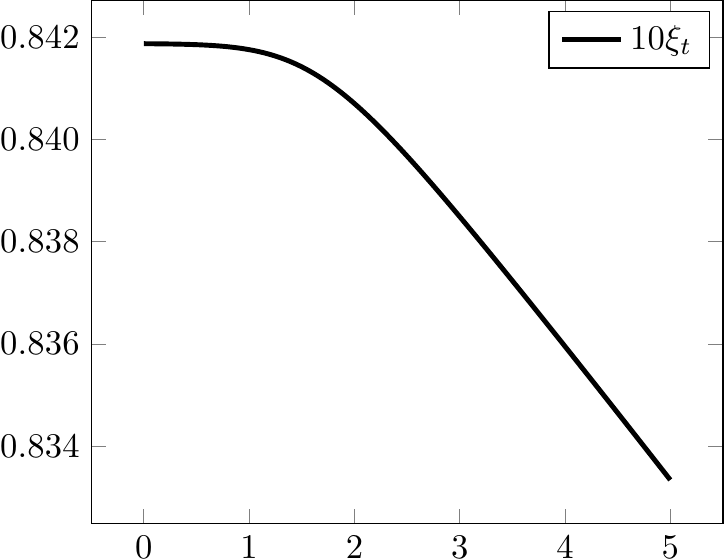} \quad
\includegraphics[width=0.40\textwidth]{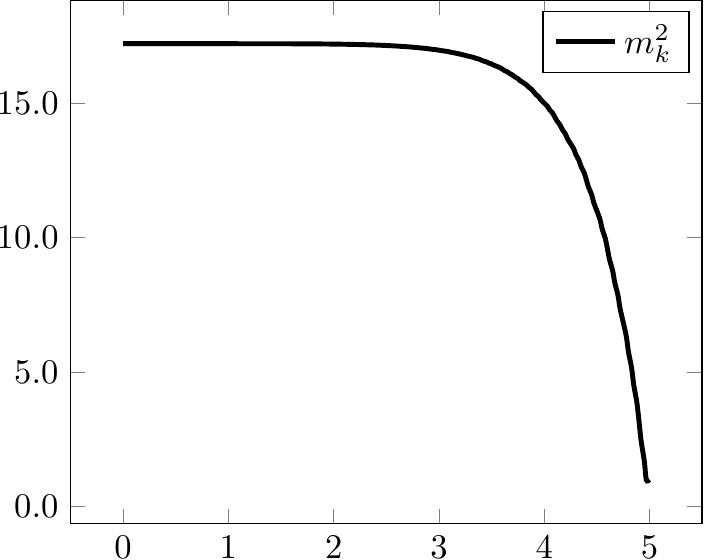}\\[3mm]
\includegraphics[width=0.39\textwidth]{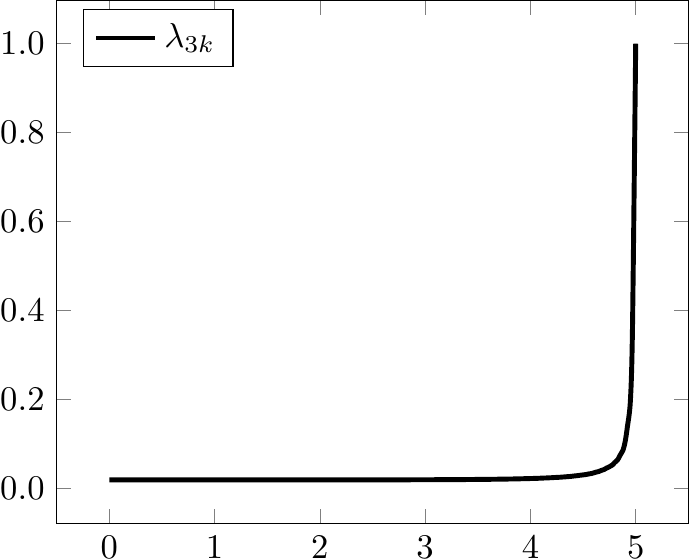} \quad
\includegraphics[width=0.41\textwidth]{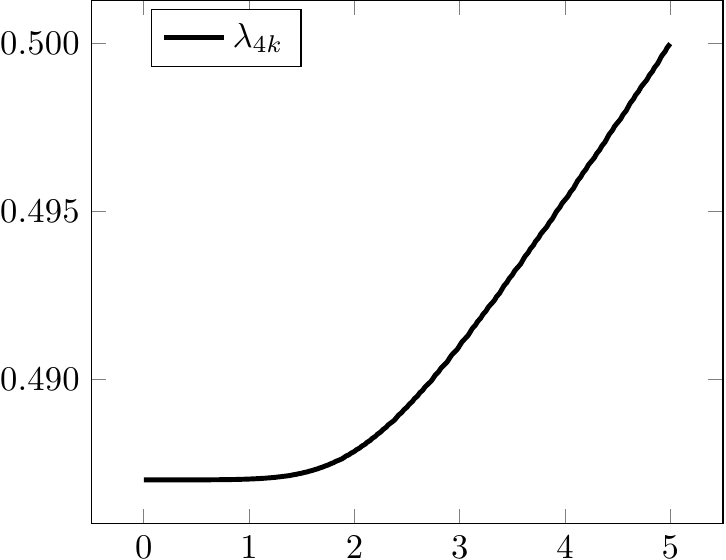}\\[3mm]
\includegraphics[width=0.41\textwidth]{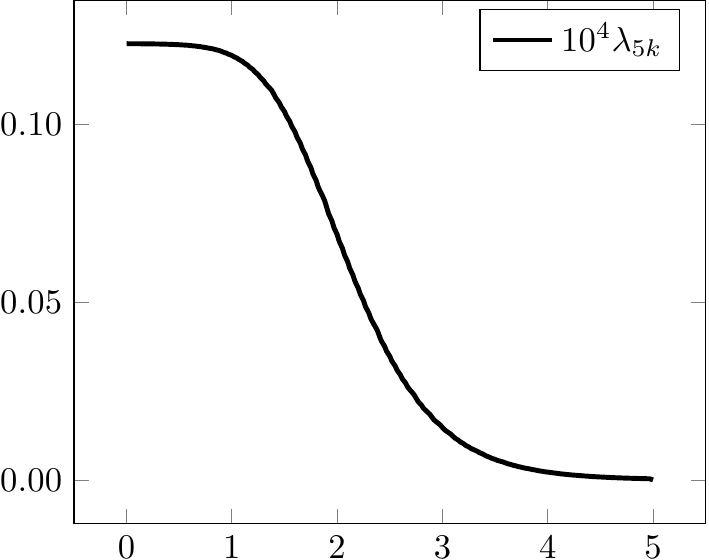} \quad
\includegraphics[width=0.40\textwidth]{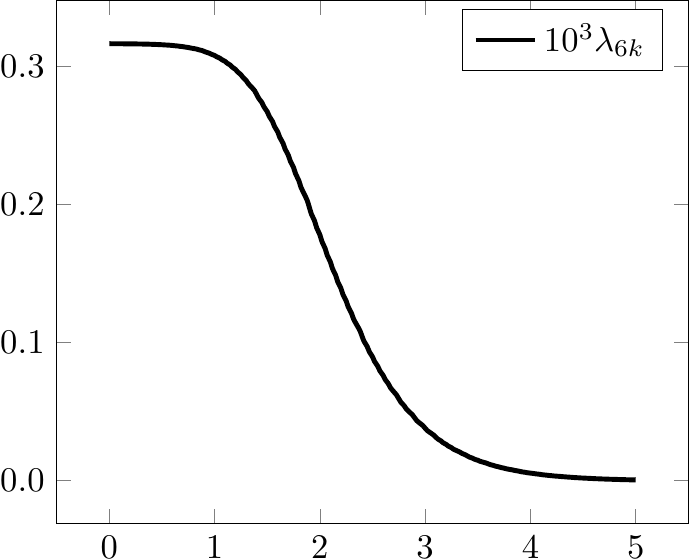}
\caption{\sl
Running of the nonminimal parameter $\xi_t$ and couplings 
$m^2_{k},\la_{3k},\dots,\la_{6k}$
(in units of $\mu$) as functions of $t=\log(k/\mu)$.
The initial data at the cut-off scale $k=\Lambda=\mu e^{5}$
are $m^2=\mu^2,\;\la_{4} = 0.5,\;\la_3=\mu,\;
\la_5=\la_{6} = 0$ and $\xi= 1/6$.}
\label{fig:parameterbroken}
\end{center}
\end{figure}

The numerical results of such an analysis are presented
at Fig. \ref{fig:phibroken} and Fig. \ref{fig:parameterbroken}.
The first plot of Fig. \ref{fig:phibroken} shows the scale-dependence
of the minimizing value of the field $\phi_{0k}$ in
the broken phase. As expected, the minimum of the potential are driven
closer to the origin by the quantum fluctuations. For the chosen
initial parameters at the cutoff $\Lambda=\mu e^{5}$
\beq
\phi_{0}\approx 3.5\mu,\quad m^2=\mu^2,\quad \lambda_3=\mu,\quad
\lambda_4=0.5,\quad \la_5=\la_6=0
\eeq
the system stays in the broken phase for all scales.
This can also be seen from the running of the couplings shown
in Fig. \ref{fig:parameterbroken}. The coupling $\lambda_{3k}$ decreases
rapidly when one moves from the UV to the IR. At the same time
the higher couplings $\lambda_{5k}$ and $\lambda_{6k}$ acquire
non-zero values, although they remain small in the IR.

We solved the flow equations with vanishing $\eta_t$.
Of course, one should choose the anomalous dimension self-consistently. But since
at the scale $k=\mu$ we have $\lambda_{3}\approx 0.0189$ and
$m^2\approx 15.2$ the first guess for the anomalous dimension
\begin{equation}
\eta_{k=\mu}\approx
\frac{1}{32\pi^2}\frac{\lambda_{3\mu}^2}{(1+m_\mu^2)^4}
\approx 1.6\cdot 10^{-11}
\end{equation}
yields a tiny value in the infrared. Thus, assuming $\eta_t=0$ is a very good
approximation.
In order to induce more dramatic qualitative changes in the
flow of the non-minimal parameter, we considered other sets of
initial parameters in the UV. For initial parameters which can
be integrated to the infrared we did not observe a strong
running of $\xi$. Thus we conclude that the qualitative form
of the flow of $\xi_t$ is not very sensitive to the initial
parameters.

%%%%%%%%%%%%%%%%%%%%%%%%%%%%%%%%%%%%%%%%%%%%%%%%%%%%%%%%%%%
%%%%%%%%%%%%%%%%%%%%%%%%%%%%%%%%%%%%%%%%%%%%%%%%%%%%%%%%%%%
\section{Conclusions}

We have constructed and explored the FRG equations for a real scalar
field in curved space-time background, including the non-minimal
parameter $\xi$ of the non-minimal interaction between scalar field
and curvature. The $\be$-functions obtained within the very simple
truncation scheme (\ref{act3}) reproduce several important features
of the standard perturbative renormalization group, including what
one can prove as being the non-perturbative universal properties of
the RG flows. First of all, the FRG equations follow the known
hierarchy of renormalization in curved space, as described by
the points {\it i) - iii)} in the Introduction. Furthermore, the
FRG trajectory for $\xi$ corresponds to the equation which is
linear in $\xi$, exactly as it should be at both perturbative
and non-perturbative levels. In the massless case, the
$\be$-function for $\xi$ has the conformal fixed point $\xi=1/6$,
which is typical for the one-loop case \cite{book}. In our opinion,
this may be the result of the restricted form of the truncation
(\ref{act3}), because higher loop corrections involve powers of
$\log(\phi)$ in both potential and kinetic sectors, which are
beyond the given approximation. It would be interesting to explore
including such terms in an FRG-analysis and we expect to do it in
a future work.

The most remarkable aspect of the RG flow for a massive theory
is a strong six-power decoupling in the IR. As a result the
running of all couplings and $\xi$ actually stops very soon
on the way from the cut-off scale in the UV down to the IR.
We conclude that the desirable strong running of $\xi$ can
not be achieved in the framework of scalar theory. At the
same time, there are chances to achieve such an effect in the
mixed theory with different mass scales, especially through
the quantum effects of relatively light or massless particles.

An interesting extension of the RG flow in the theory with the
non-minimal parameter is related to the broken phase, when one can
also observe the wave-function renormalization and its effect on
the RG trajectories for couplings and $\xi$. We have found that,
regardless of the more complicated form of the RG flow, the
qualitative form of the flow for $\xi$ remains the same, in the
sense that the numerical effect of the scale-dependence is quite
small in the scalar theory. One of the consequences is that the
scale-dependence in the non-local part of the induced action of
gravity in the theory with Spontaneous Symmetry Breaking will be
also small. However, as it was discussed in \cite{PoImpo}, any
form of scale-dependence may have a significant impact on the
induced cosmological constant term and especially on its
non-local extensions. Therefore, the problem of the running of
$\xi$ from UV to IR deserves further detailed studies, especially
in more general theories which involve several mass scales.

%%%%%%%%%%%%%%%%%%%%%%%%%%%%%%%%%%%%%%%%%%%%%%%%%%%%%%%%%%%
%%%%%%%%%%%%%%%%%%%%%%%%%%%%%%%%%%%%%%%%%%%%%%%%%%%%%%%%%%%
%%%%%%%%%%%%%%%%%%%%%%%%%%%%%%%%%%%%%%%%%%%%%%%%%%%%%%%%%%%
\section*{Acknowledgements}
I. Shapiro is grateful to the Theoretisch-Physikalisches-Institut
of the Friedrich-Schiller-Universit${\ddot {\rm a}}$t in Jena for
warm hospitality and to the University of Geneva, CNPq, FAPEMIG and
ICTP for partial support. P. Teixeira is grateful to CAPES for
supporting her visit to SISSA, and to the Theoretical Particle
Physics of SISSA for warm hospitality. A. Wipf thanks
Omar Zanusso for fruitful discussions and the DFG
for supporting this work under grant no. Wi777/11-1.

%%%%%%%%%%%%%%%%%%%%%%%%%%%%%%%%%%%%%%%%%%%%%%%%%%%%%%%%%%%

\end{document}